# A Syntactic Classification based Web Page Ranking Algorithm

**Debajyoti Mukhopadhyay** [1,3], **Pradipta Biswas** [2], **Young-Chon Kim** [3]

[1] Web Intelligence & Distributed Computing Research Lab, Techno India (West Bengal University of Technology)
EM 4/1 Salt Lake Sector V, Calcutta 700091, India; Email: debajyoti.mukhopadhyay@gmail.com

[2] Indian Institute of Technology, School of Information Technology
Kharagpur 721302, India; Email: pbiswas@sit.iitkgp.ernet.in

[3] Chonbuk National University, Division of Electronics & Information Engineering
561-756 Jeonju, Republic of Korea; Email: yckim@chonbuk.ac.kr

**Abstract.** The existing search engines sometimes give unsatisfactory search result for lack of any categorization of search result. If there is some means to know the preference of user about the search result and rank pages according to that preference, the result will be more useful and accurate to the user. In the present paper a web page ranking algorithm is being proposed based on syntactic classification of web pages. Syntactic Classification does not bother about the meaning of the content of a web page. The proposed approach mainly consists of three steps: select some properties of web pages based on user's demand, measure them, and give different weightage to each property during ranking for different types of pages. The existence of syntactic classification is supported by running fuzzy c-means algorithm and neural network classification on a set of web pages. The change in ranking for difference in type of pages but for same query string is also being demonstrated.

## 1 Introduction

The World Wide Web is an architectural framework for accessing linked documents spread out over millions of machines all over the Internet. It began in 1989 at CERN, the European center of Nuclear Research. At that time FTP data transfers accounted for approximately one third of Internet traffic, more than any other application. But after the introduction of WWW it had a much higher growth rate. By 1995, Web traffic overtook FTP data transfer and by 2000 it overshadowed all other applications. The popularity of WWW is largely dependent on the search engines. Search engines are the gateways to the huge information repository at the internet. Now anyone can quickly search for helpful cleaning tips, music lyrics, recipes, pictures, celebrity websites and more with search engines. Search engines consist of four discrete software components: Crawler or Spider: a robotic browser like program that downloads web pages; Indexer: a blender like program that dissects web pages that are downloaded by spiders; The database: a warehouse of the pages downloaded and processed; Search engine results engine: digs search results out of the database.

The web page ranking algorithms play their role at the last component. Exactly what information the user wants is unpredictable. So the web page ranking algorithms are designed to anticipate the user requirements from various static (e.g., number of hyperlinks, textual content) and dynamic (e.g., popularity) features. They are important factors for making one search engine better than another. A short paper introducing an algorithm called FlexiRank has already been published in 2005 [18]. A detailed description of this research work is presented in this paper. This algorithm offers some flexibility to the user while searching the web pages with a proposed search engine interface through a textbox with some option buttons to fine-tune the options while sending the query to the search engine. The option buttons are easy to use for naïve users and not as complicated as some of the existing advanced search engine interfaces.

## 2 Related Work

Among the existing page ranking algorithms the most important algorithms are Kleinberg's HITS algorithm, Brin & Page's PageRank algorithm, SALSA algorithm, CLEVER Project etc. The AltaVista Search Engine implements HITS algorithm [1]. But the HITS (Hyperlink Induced Topic Search) is a purely link structure-based computation, ignoring the textual content [2][3]. According to PageRank algorithm used in Google [4], a page has a high rank if the sum of the ranks of its back-links is high. The benefits of Google PageRank are the greatest for under specified queries, for example: 'Stanford University' query using PageRank lists the university home page first. CLEVER project [5] focuses on higher-level applications based on the basic CLEVER engine. It mainly emphasizes on Enhancements to HITS algorithm, hypertext classification, focused crawling, mining communities, modeling the web as a graph. Actually this system ranks the pages primarily by measuring links between them. It assigns to each link a non-negative weight which depends on the query term and end point. The main difference between Google PageRank method and Hub/Authority (CLEVER, C-Server) page ranking methods is that Google PageRank assigns initial ranking and retains them independently from queries (so it is fast) whereas the Hub/Authority methods assemble different root set and prioritizes pages in the context of query [3]. Companion Algorithms introduce a new concept which is exploiting not only links but also their order on a page [3]. The weight assignment to hyperlinks is more exploited in [6] where each link gets a weight based on its position at the page, length of anchor text and on the tag where the link is inserted. In [7] the links of a web page are weighted based on the number of inlinks and outlinks of their reference pages. The resulting algorithm is named as 'weighted page rank'. These two page ranking algorithms [6] [7] does not take any extra information from the surfer for giving an accurate ranking. In [8] a new approach of dissecting queries into crisp and fuzzy part has been introduced. The user interface is proposed to be divided into two phase. The first phase will take crisp queries whereas the second phase consider the fuzzy part (like the words popular, moderate distance etc.) of the query. Efforts are also been taken to make the ranking more accurate by incorporating topic preference of user during ranking. In [9], a parameter viz. query sensitiveness is measured which signifies the relevance of a document with respect to a term or topic. The scope of search engine is divided into global and local scope. The local scope is developed from inverse document table and used to measure the query sensitiveness of a page. The pages are ranked based on two parameters-their global importance and query sensitiveness. In [10], the damping factor of page rank algorithm is changed to a parameter viz. confidence of a page with respect to a particular topic. The confidence is defined as the probability of accessing a particular page for a particular topic.

## 3 Existing Problem and Our Approach

Generally an Internet surfer does not bother to go through more than 10 to 20 pages shown by a search engine. So the web page ranking should concentrate very much for giving higher rank to the relevant pages. In spite of all sophistication of the existing search engine, sometimes they do not give satisfactory result [6] [7][8]. The reason is that most of the time a surfer wants a particular type of page like an index page to get the links to good web pages or an article to know details about a topic. What is lacking in the existing search engines is a proper classification of the search pages and ranking according to that. The advanced search options of search engines take very raw input like link to or from a particular website, or mandatory portion of a query etc, which are easier to use for a search engine software but difficult to interpret and use for a layman or non computer professional person. A lot of work has been done on web classification but that are based on semantics of the content of pages [11]. In [12] the classification is based on co-citations among web pages. But the existing classifiers are like Education, Art & Humanity, and Entertainment etc. This type of classification is of not much help to web page ranking for most of the search strings. For example if a search topic like "Human Computer Interaction" is given, it is easy to guess that education related pages are wanted; there is no need of using any extra knowledge to derive the user's demand for the proper class of pages. In [8] the user's intention is felt by identifying and separately analyzing the fuzzy portion of a query. The changed interface of a search engine in this case will not be much user friendly. The approaches taken in [9] or [10] do not take the user preference explicitly. So the ranking has no control over a particular type of page.

Approach taken in this paper is to make a classification of web pages based on only syntax of the page and change the search engine interface to take the proper class of a page alongwith the query topic. The web page classification will be like Index page, Home Page, Article, Definition, Advertisement Pages etc. This type of classification is independent to the semantics of the content of a page, so it can provide useful information for ranking a page according to the user's demand. As for example, if a search topic is given like "Antivirus Software" and given category of page is "Homepage" then the homepages of different Antivirus companies will get higher rank. If for the same query, the category given is "Article", then the pages giving general description of Antivirus Software will get higher rank. Again if the given category is "Index" then a page having large number of links to different antivirus software vendors will get higher rank. So for a single query term a particular page can get different ranking based on user's demand. In the proposed page-ranking algorithm some properties of a web page are identified which are blind to the meaning of the content of a page. These properties are parameterized and measured separately from each other. The page rank is calculated by taking a weighted average of different parameters. The weight assigned to each parameter depends upon the category of the page wanted by the user.

## 4 Parameters Used for Ranking

In this section different parameters, selected for web page ranking, are discussed. The page ranking will be done by taking a weighted average of all or some of the parameters. The weight given to a particular parameter will depend upon the category of the page. In the proposed algorithm a single query may give different ranking to a page depending on the category of the page-which is not possible in any existing search engines. The algorithm is flexible in the sense that just by changing the weights the same algorithm provides ranking for different types of pages.

### 4.1 Relevance Weight

Relevance weight measures the relevance of a page with respect to a query topic by counting the number of occurrences of the query topic or part of the query topic within the text of the document. The term frequency matrix provides useful information for calculating relevance weight. Some existing ways are Vector Space Model [15] [16], Cover Density Ranking [17], Three Level Scoring method [14] etc.

In the present paper the page relevance algorithm used, has taken an approach of the Three Level Scoring method. The relevance of a page is gradually increased as more portion of the query topic occurs in the page. As for example a search string like "Data Mining" may be considered. Let in Page A the string "Data Mining" appears. In Page B only the word "Mining" appears. Now it may happen that Page B refers to "Coal Mining" which is not at all related to the search string "Data Mining". So it can be inferred that, in a web page where the entire search string appears as a whole is more relevant to the search topic than a page where only part of the string appears.

In the proposed algorithm the words in "Stop List" are removed first from the search string. After proper stemming, the relevant keywords or terms are extracted from the search string. Next, the occurrence of each keyword and term are found out, and a weightage is given to it as the ratio of its length to the length of the given query topic. So for the above example the term "Data Mining" will get a weightage of 1 whereas the term "Mining" will get a weightage of "6/11" i.e. 0.545. Finally the algorithm is as follows:

```
function Calc_Relevance_Wt(File F: A Text File, String S: The Search
String)

return Relevance_weight /* relevance of textual content of file F w.r.t. Search
string S */
var KEYWORD_SET[1…N]   /* To store the subset of relevant strings within the search
string */
var CNT /*Number of relevant substrings */
var OCCURRENCE[1…N]   /* OCCURRENCE[I]= Occurrences of substring KEYWORD_SET[I]
within file F */

KEYWORD_SET=Set of relevant substrings within S
CNT=|KEYWORD_SET|
```

```
For (I=1 to CNT)
     OCCURRENCE[I]= Number of Occurrences of substring KEYWORD_SET[I] within
file F

For (I=1 to CNT)
     Relevance_Weight=Relevance_Weight+(Length(KEYWORD_SET[I])/Length(S))*
OCCURRENCE[I]
```

### 4.2 Hub and Authority Weight

Hub and authority weight of a page is calculated using the HITS algorithm. Given a user query, the HITS algorithm first creates a neighborhood graph for the query. The neighborhood contained nearly top 200 matched web pages retrieved from a content-based web search engine; it also contained all the pages these 200 web pages linked to and pages that linked to these 200 top pages.

Then, an iterative calculation was performed on the value of authority and value of hub. For each page p, the authority and hub values are computed as follows:

$$A_p \leftarrow \sum_{q \in L(p)} H_q \qquad H_p \leftarrow \sum_{q \in L(p)} A_q$$

The authority value of page p is the sum of hub scores of all the pages that points to p, the hub value of page p is the sum of authority scores of all the pages that p points to (Fig.1). Iteration proceeded on the neighborhood graph until the values converged.

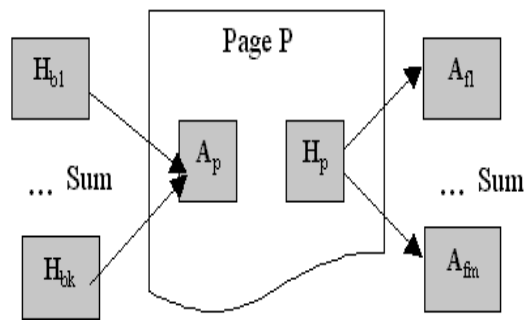

**Fig.1** Hub and Authority Weight of a Page

It has been claimed that the small number of pages with the largest authority converged value should be the pages that had the best authorities for the topic. And the experimental results support the concept.

### 4.3 Link Analysis of a Page

The HITS algorithm analyzes the link structure information of a web graph. The hyperlink information of a single page (e.g. number of links, anchor text and positions of the pages in the domain tree with respect to a particular page) are also found to give useful information during syntactic categorization of a web page.

#### 4.3.1 Number of Hyperlinks

The number of hyperlinks of a page is calculated by getting the total number of *a href* tags. For getting the exact the number of hyperlinks the number of *frame src* tags should be added to the number of *a href* tags and links to the same page should be excluded.

#### 4.3.2 Anchor Text

The anchor text can be used to calculate the weight of links during measuring hub and authority weight. By analyzing anchor text the glossary pages can very easily be identified.

**4.3.3 Positions of Hyperlinked Pages in the Domain Tree with Respect to a Particular Page**

It has been found the portals have large number of hyperlinks pointing to same level nodes in the domain tree rooted at the next higher level node of the source of the page; e.g., if source is **a.b.com** nature of hyperlinks are **x.b.com** or **y.b.com**. The site maps and home pages have large number of hyperlinks pointing to lower level nodes in the domain tree rooted at the source of the page; e.g., if source is **a.b.com** nature of hyperlinks are **a.b.com/x**, **a.b.com/y**.

**4.4 Types of Content**

The syntactic analysis of the content also gives useful properties about the type of a page. Examples of these types of properties are:

1. Number of images in a page
2. Text length to number of images proportion etc.
3. Relevance weight of the query string within special tags like Heading tag, title tag etc.

## 5 Properties of Different Types of Pages

Here, different types of pages and their unique properties are identified. The result is shown in Table 1. The list is not an exhaustive one; still the identified properties can be used to develop the criteria or properties for syntactic classification. They are also used to develop heuristics during web page classification.

## 6 The Algorithm

This algorithm operates on a set of web pages returned by a web crawler and gives a ranking of the pages as output. It operates according to the following steps.

- **Select attributes based on user demand:** Based on the user's demand the algorithm chooses a set of properties of a web page. Some properties are chosen irrespective of the user's demand. Examples of such mandatory properties are Relevance weight, Hub weight and Authority weight. The other attributes are chosen based on user demand to provide an accurate ranking. Examples of such optional attributes are number of hyperlinks, number of images etc.

- **Measure the attributes:** The selected attributes are measured for each web page.

- **Calculate rank:** The rank is calculated by taking a weighted average of the measured values. The weight assigned to each attribute is based on user's demand.

The algorithm provides flexibility in two grounds:

- **In selection of properties:** As for example when the user's demand is Index type pages, number of hyperlinks of a page will be measured whereas number of images or text to image proportion will not be measured.
- **In determining weight of properties:** The selected attributes get different weights for difference in user demand. As for example, for article type of pages relevance weight and authority weight will get highest weightage whereas for advertisement type of pages number of thumbnails (i.e., number of images) and hub weight will get higher weightage.

Due to these varying selections of properties and their corresponding weightages, the algorithm provides more flexibility to the user and also gives more accurate result.

**Table 1**. Different types of Pages and their Characteristics

| Sl No | Type of Pages | Characteristics |
|---|---|---|
| 1 | Index | 1. Having small number of hyperlinks normalized w.r.t. document length<br>2. Having large hub weight |
| 2 | Homepage | 1. Having large number of hyperlinks pointing to lower level nodes in the domain tree rooted at the source of the page<br><br>e.g., if source is a.b.com, nature of hyperlinks are a.b.com/x, a.b.com/y…..<br><br>2. Having a few images |
| 3 | Portal | 1. Having large number of hyperlinks pointing to same level nodes in the domain tree rooted at the next higher level node of the source of the page<br><br>e.g., if source is a.b.com nature of hyperlinks are x.b.com, y.b.com ….. |
| 4 | Article | 1. Have large Document Length<br>2. Having small number of hyperlinks normalized w.r.t. document length<br>3. High relevance weight |
| 5 | Advertisement Pages | 1. Large number of thumbnails<br>2. Large number of links to dynamic web pages |
| 6 | Research papers | 1. Very few hyperlinks<br>2. Large length<br>3. Mostly are in .ps or .pdf format |
| 7 | Glossary | 1. Anchor texts like A, B, C,….,Z |
| 8 | Tutorial | 1. Have large Document Length<br>2. Having small number of hyperlinks normalized w.r.t. document length |
| 9 | Definition | 1. High relevance weight normalized w.r.t. length of the document |
| 10 | Downloads | 1. Presence of links to .zip, .tar, .gz, .exe |

## 7 Experimental Result

The experiment has been done in two parts. In the first part several web pages are downloaded and classified according to the proposed properties. In the second part some web pages are downloaded again from an existing search engine and ranked according to this algorithm. Each of these parts is discussed below.

### 7.1 Clustering Web Pages

In this part about 50 web pages are downloaded from Google search engine. The pages are clustered according to the following properties:
1. Relevance weight
2. Number of Images
3. Number of Links
4. Number of Self Links
5. Number of links to same level pages in the domain tree (refer Sec 4.3.3)
6. Number of links to lower level pages in domain tree (refer Sec 4.3.3)
7. Document Length

For clustering purpose Fuzzy c-means [11] algorithm was used. Cluster validation is done by Classification Entropy. With c = 4 clear classification has been got between pages of short length and long length and text intensive pages (article or paper type pages) and hyperlink intensive pages (index type page). Properties of the cluster centers are shown in Table 2.

**Table 2:** Cluster Centers

| Cluster id | Relevance Weight | Number of Images | Number of Hyperlinks | Number of Lower Level Links | Number of Same Level Links | Document Length |
|---|---|---|---|---|---|---|
| 1 | 13.238 | 13.154 | 12.045 | 1.963 | 4.199 | 26213.667 |
| 2 | 8.587 | 6.836 | 55.507 | 8.618 | 10.968 | 17108.8 |
| 3 | 3.536 | 15.78 | 27.089 | 6.3 | 10.193 | 16262.286 |
| 4 | 21.295 | 3.714 | 34.64 | 9.166 | 18.041 | 22890.75 |

As can be seen in Table 2, Cluster 1 and 4 stand for pages with high relevance weight. Cluster 2 stands for pages with large number of hyperlinks and of short length. The difference between same level and lower level links are greater for cluster 1 and 4. So it can be inferred web pages belong to cluster 1 and 4 offer some services or portals type of pages.

Clustering is an example of unsupervised learning. From the clustering result we can get a hint of the existence of syntactic classification, but it is not yet confirmed. So we go for a classification of web documents according to the previously mentioned criteria using a neural network. A software viz. NeuNet Pro, which is a tool for pattern recognition, data mining, modeling and prediction using neural networks, has been downloaded from the site of CorMac Technologies Inc (http://www.cormactech.com). Using this software we have defined a feedforward neural network with 5 hidden nodes and use back-propagation learning algorithm for classifying 30 web pages downloaded from Google. After completing 1000 cycles with learning rate=60 and verify rate=10 (these rates are defined by the software internally) we get the following scatter graph and time series graph as shown in Fig. 2 and Fig. 3 respectively.

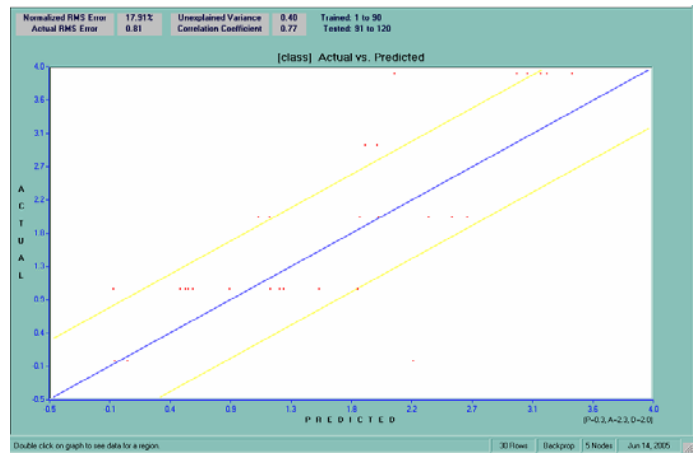

**Fig. 2** Scatter Graph for Syntactic Classification

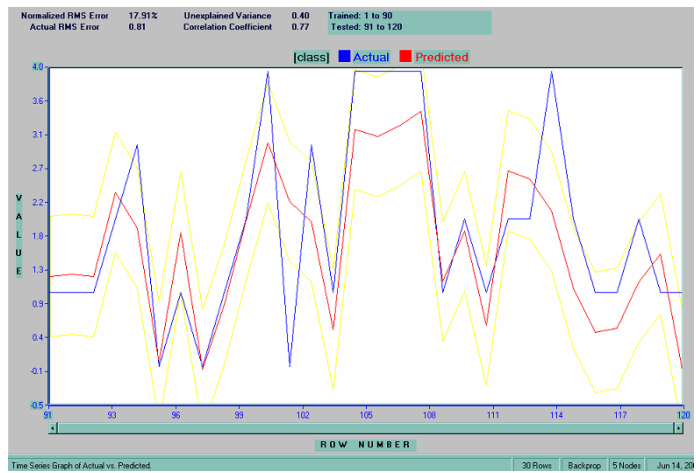

**Fig. 3** Time Series Graph for Syntactic Classification

Since the classification is carried on using only 7 properties, we did not get a very accurate classification. Still the result of the fuzzy clustering algorithm and the less than 20% R.M.S. error in classification confirm the existence of syntactic classification of web pages.

### 7.2 Ranking the Web Pages

For testing the actual change in ranking for different types of pages, the proposed ranking algorithm is run on top 30 pages downloaded using Google search engine with the search topic "Human Computer Interaction". The screenshot of the proposed interface of a search engine is shown in Fig 4.

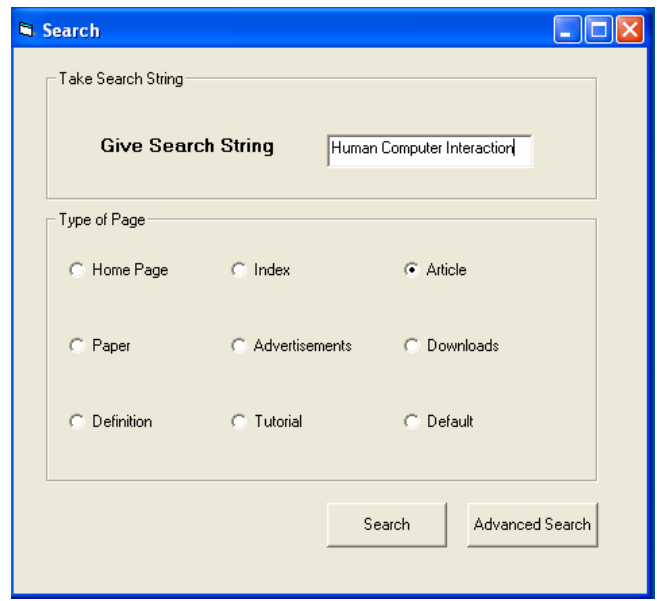

**Fig 4.** Screenshot of the Proposed Interface of a Search Engine

When the type of page is given as index the following three pages get first three ranks.
1. http://is.twi.tudelft.nl/hci/
2. http://dmoz.org/Computers/Human-Computer_Interaction/
3. http://www-hcid.soi.city.ac.uk/

The first two pages are literally index pages while the third one is the home page of Centre of HCI Design, City University London. The page contains a lot of hyperlinks.

Again when the type of page is given as article the following three pages get first three ranks.
1. http://sigchi.org/cdg/cdg2.html
2. http:// www.cs.cmu.edu/~amulet/papers/uihistory.tr.html
3. www.id-book.com/

Here also the first two sites are text intensive articles.

As can be seen in the interface a default option is also being kept for ranking in all types of pages.

## 8 Work in Progress

Currently the system is working fine on 9 to 10 types of pages. But the vast WWW can not be classified properly using only 9 to 10 categories. There also exists some interdependency among different classes of pages. We are trying to develop an ontological structure on the types of pages. The classifier being developed will be an *Adaptive Neural Fuzzy Inference System (ANFIS)*. Effort is being made to enlarge the training data set in order to make the weights more accurate.

## 9 Conclusion

The present paper discusses a web page ranking algorithm, which consolidates web page classification with web page ranking to offer flexibility to the user as well as to produce more accurate search result. The classification is done based on several properties of a web page which are not dependent on the meaning of its content. The existence of this type of classification is supported by applying fuzzy c-means algorithm and neural network classification on a set of web pages. The typical interface of a web search engine is proposed to change to a more flexible interface which can take the type of the web page along with the search string.